\documentstyle[12pt]{article}
\begin{document}
\begin{titlepage}
\thispagestyle{empty}

\vspace*{2cm}

\begin{center}
{\Large \bf Meson--exchange contributions \\ to the \\[.2cm]
nuclear charge operator}

\vspace{1.5cm}
{\large Antonio M. Lallena}

\vspace{.3cm}
{Departamento de F\'{\i}sica
Moderna, Universidad de Granada, \\
E-18071 Granada, Spain}

\end{center}

\vspace{2cm}
\begin{abstract}
The role of the meson--exchange current correction to the
nuclear charge operator is studied in electron scattering
processes involving the excitation of medium and heavy nuclei to
energies up to the quasi--elastic peak.  The effect of these
contributions in the quasi--free electron scattering process is a
reduction of at most a 3\% in the longitudinal response at the
energy of the peak, a value which is below the experimental
error and must not be taken into account in calculations in this 
energy region. On the other hand, the excitation of low--lying nuclear
levels of neutronic character shows, with respect to the
protonic ones, a considerable effect due to the inclusion of the 
two--body term in the charge operator. More realistic
calculations, such as those performed in the random--phase
approximation framework, give rise to a mixing of one
particle--one hole configurations of both kinds which reduce
these effects. However, it has been found that the excitation of 
some of these levels is sizeably affected by the meson--exchange 
contribution. More
precise experimental data concerning some of these states, such
as e.g. the high--spin states in $^{208}$Pb, could throw some
light in the problem of a more feasible determination of these effects
and, as a consequence, could provide an alternative procedure to
obtain the charge neutron form factor.
\end{abstract}

\end{titlepage}

\newpage

\setcounter{page}{1}

\section{Introduction}

The exchange of mesons between nucleons in a nucleus, which is
the mechanism responsible for the nucleon--nucleon interaction,
modifies the electromagnetic interactions of the nucleus in
order to maintain the current conservation. This obvious
statement was pointed out by Siegert~\cite{Sie37} soon after the
exchange model of the nuclear forces was proposed by
Yukawa~\cite{Yuk35}. The realization that this exchange could
produce non--negligible effects on nuclear observables, such as
the nuclear magnetic moments, preceeded the discovery of the
pion~\cite{Vil47}. 

Nevertheless, it was in the early 70's when this topic started to
be of general interest. The reason for that lied in the
increasing amount of experimental data of enough accuracy which
begun to be available at that time.

The first clear evidence of the importance of the so--called
meson--exchange currents (MEC) was identified in connection with
the radiative thermal neutron capture reaction by a
proton~\cite{Ris72}. After this, processes such as the deuteron
electrodisintegration at threshold~\cite{Hoc73} and the
radiative neutron capture by the deuteron~\cite{Had73} and
observables such as the magnetic form factors of $^3$H and
$^3$He~\cite{Bar75} could be explained both qualitative and 
quantitatively only after the consideration of MEC contributions.

Most of the work done since that time has been devoted to the
investigation of the contributions of MEC to electromagnetic
processes. These contributions have been studied in a wide
range of momentum transfers, from the ground state to excitation
energies corresponding the quasi--elastic (QE) peak and beyond, for
light, medium and heavy nuclei and using quite different models
to describe the nuclear structure. Sizeable effects have been
found in a number of situations, mainly in $A\leq 4$ nuclei. In
medium and heavy nuclei, MEC produce rather small effects 
what has been linked to the absence of short--range correlations 
of tensor type in the models commonly used to describe 
the wave functions of these nuclei.~\cite{Ang96,Lei90}

The reason of the focussing on multipole transitions involving
the nuclear current only is related to the Siegert
theorem~\cite{Sie37} which establish that, at low momentum
transfer, charge form factors are insensitive to meson exchange effects
and that they can be reasonably well described by the single
nucleon (impulse) approximation. However, Siegert theorem only
applies in the long wavelength limit and it is possible to find
situations in which also these form factors are modified by the
presence of exchange effects.

In principle, and this can also be applied to transverse form
factors, meson exchange contributions will be relatively bigger
the larger the momentum transfer is. However, at high-$q$ other
effects such as relativistic effects or short--range
correlations enter into play and make difficult a clear
discrimination of the different contributions. Therefore the
only way to study meson exchange effects is to look at some 
relatively small value of $q$ where the
contributions of the single--nucleon operators vanish
accidentally. This is what
happens in some of the processes above mentioned and this can
also occur, and in fact it does, for charge form factors even at 
relatively low momentum transfer.

The first time that pion--exchange was recognized as responsible
of a sizeable effect in nuclear charge form factors was in 1974
when Kloet and Tjon~\cite{Klo74} calculated the charge form
factors of the $^3$H and $^3$He. Basically it was impossible to
atteint the agreement with the experimental data in the second
maximum using single--nucleon charge operator, even with
relativistic wave functions. Additional calculations in light
nuclei such as the deuteron~\cite{Jac75} and $^4$He~\cite{Bor75}
showed similar results.

In heavier nuclei the situation is more complicated because of
the incertitudes due to the nuclear structure problem. The
presence of many single--particle orbitals hides to some extent
the signature of MEC effects and one should be
careful with the conclusions drawn in this direction. In any
case some calculations have been done in which those
contributions are investigated. Radomski and Riska~\cite{Rad76}
have evaluated the pion--exchange effect in nuclear charge form
factors and charge distributions of $^{16}$O and $^{40}$Ca and
have found smaller contributions than for the
$\alpha$--particle. On the other hand,
Negele and Riska~\cite{Neg78} have shown that the inclusion of the
pion--exchange term in the charge operator brings calculated
charge form factors into better agreement with the experiment
for closed--shell nuclei troughout the periodic table. Riska
and Str\"uve~\cite{Ris83} have calculated the charge form
factors again for the same nuclei, paying attention to
the role of MEC contributions in connection with medium polarization
corrections and short--range effects. Finally, Lodhi and 
Hamilton~\cite{Lod85} have investigated the charge form factor
of $^6$Li by considering MEC and short--range correlations,
simultaneously. Despite the fact that in
Refs.~\cite{Rad76,Ris83,Lod85} a simple harmonic oscillator
shell model has been used, the results can be considered as
feasible because no important differences are found when these
results are compared with those obtained with
Hartree--Fock~\cite{Neg78} or
Brueckner--Hartree--Fock~\cite{Gar76} wave functions.

In any case, not much work has been done in this context. In
this paper, we want to investigate the MEC contributions to the 
nuclear charge operator in a variety of situations involving
electromagnetic excitations of medium and heavy nuclei. The main
purpose is to understand how such contributions affect the
results one obtains by means of the impulse approximation (IA),
which is the one widely
admitted to calculate the different observables in this sector.

The organization of the paper is as follows. In Sec.~II we
discuss the details concerning the charge operator and the
corrections to be considered. Sec.~III is devoted to analyze the
modification produced by the new operators in the nuclear
response in the QE peak. In Sec.~IV we study in detail the
results obtained for the electroexcitation of bound levels 
in closed--shell nuclei both in a shell--model approach and
in a more realistic calculation performed in the framework
of the random--phase approximation (RPA).
We finish by summarizing the results and giving our conclusions
in the last section.

\section{Model for the nuclear charge operator}

Our purpose is to calculate different observables corresponding
to the electron scattering by nuclei, such as cross--sections,
form factors, response functions, etc. To do that it is
necessary to fix both the model for the electromagnetic operator
and the nuclear structure approach to be considered to describe
the nuclear states. In this section we focus our attention in
the first one.

The larger contributions to the observables mentioned above are
produced by the individual nucleons. In this approximation, the
IA, the nuclear charge operator is
\begin{equation} \label{ch}
\rho^{\rm IA}({\bf q},\omega) = 
\sum_{k=1}^{A}  \left [ G_{\rm E}^{\rm P}({\bf q},\omega) 
\frac{1+{\tau}_3^{k}}{2} + G_{\rm E}^{\rm N}({\bf q},\omega)
\frac{1-{\tau}_3^{k}}{2} \right] , 
\end{equation}
while the nuclear current 
\[
{\bf j}^{\rm IA}({\bf q},\omega)={\bf j}^{\rm C}({\bf q},\omega)+
{\bf j}^{\rm M}({\bf q},\omega)
\]
includes two terms, the convection current
\[
{\bf j}^{\rm C}({\bf q},\omega)=\sum_{k=1}^A  \left
( G_{\rm E}^{\rm P}({\bf q},\omega) \frac{1+{\tau}_3^{k}}{2} + 
  G_{\rm E}^{\rm N}({\bf q},\omega)
\frac{1-{\tau}_3^{k}}{2} \right ) \left ( \frac{{\bf p}_k  + 
{\bf p}'_k}{2M_k} \right )
\]
and the magnetization current
\[
{\bf j}^{\rm M}({\bf q}, \omega)=\sum_{k=1}^A \frac{-i}{2M_k}
\left ( G_{\rm M}^{\rm P}({\bf q},\omega)
\frac{1+{\tau}_3^{k}}{2} + G_{\rm M}^{\rm N}({\bf q},\omega)
\frac{1-{\tau}_3^{k}}{2} \right ) {\bf q} \times 
{\mbox{\boldmath $\sigma$}}^k .
\]
In these equations, $k$ runs over all the nucleons in the nucleus,
$G_{\rm E}^{\rm P}$ and $G_{\rm E}^{\rm N}$ 
($G_{\rm M}^{\rm P}$ and $G_{\rm M}^{\rm N}$) are the electric (magnetic)
form factors of the proton and the neutron, ${\bf p}_k$ y ${\bf p}'_k$
are the initial and final momenta of the $k$-th nucleon, $M_k$
is its mass and ${\bf q}$ and $\omega$ are the momentum and
energy transferred to the nucleus in the process.

The necessity of including MEC in the nuclear current operator
can be understood, in a very simple way, by means of the 
non--relativistic continuity equation (CE) which,
in coordinate space, reads
\[
\nabla \cdot {\bf j} = - i [H,\rho]
\]
Taking into account that
\[
\nabla \cdot {\bf j}^{\rm IA} \equiv 
\nabla \cdot {\bf j}^{\rm C} \equiv 
= - i [T,\rho^{\rm IA}]
\]
with $T$ the kinetic energy operator, if $H$  
includes a two--body potential, $V$, an
additional term in the current, ${\bf j}^{\rm MEC}$, verifying
\[
\nabla \cdot {\bf j}^{\rm MEC} 
= - i [V,\rho^{\rm IA}]
\]
must be included in order to maintain the CE.
However, this equation does not provide an unique current for
$H$ and $\rho$ given. In effect all the currents of the form 
${\bf j}+\nabla \times {\bf \eta}$ satisfy the CE independently
of the form of ${\bf \eta}$, and this happens even at the level of
the IA.

Following the nomenclature of Riska~\cite{Ris83a} we call {\it model
independent\/} current operators to those fixed by the CE. On the
contrary, those currents not affected by this equation are
referred as {\it model dependent}. Out of the first type one can
note the so--called seagull current,
\[
{\bf j}^{\rm S}({\bf q},\omega)   =  
\sum_{k<j} \left \{-i  4\pi  
\frac{f_\pi^2}{\mu^2}  F_{\rm S}({\bf q},\omega) 
[{\mbox{\boldmath $\tau$}}^k\times {\mbox{\boldmath
$\tau$}}^j]_3 
{\mbox{\boldmath $\sigma$}}^k  
\frac{{\mbox{\boldmath $\sigma$}}^j \cdot 
{\rm {\bf q}}_j}{{\mbox q}^2_j
 +  \mu^2 - \epsilon_j^2 }   \nonumber \\
 + \; \; (k \longleftrightarrow j) \right \} ,
\]
and pionic current,
\[
{\bf j}^{\pi}({\bf q},\omega) =
\sum_{k<j}  i  4\pi  \frac{f_\pi^2}{\mu^2}  F_{\pi}({\bf q},\omega)
[{\mbox{\boldmath $\tau$}}^k\times {\mbox{\boldmath
$\tau$}}^j]_3 \left ({\rm {\bf q}}_k  -  
{\rm {\bf q}}_j \right )
\frac{{\mbox{\boldmath $\sigma$}}^k \cdot 
{\rm {\bf q}}_k}{{\mbox q}_k^2 +
 \mu^2 - \epsilon_k^2 }  
\frac{{\mbox{\boldmath $\sigma$}}^j \cdot
 {\rm {\bf q}}_j}{{\mbox q}_j^2 +
 \mu^2 - \epsilon_j^2 } .  
\]

Between the model dependent ones the most relevant is the isobar
current
\begin{eqnarray*}
{\bf j}^\Delta({\bf q},\omega) & =  &
\sum_{k<j} \left [ 
\rule[-2ex]{0ex}{4ex}
 -i   4 \pi  \frac{f_\pi^2}{\mu^2}  F_{\Delta}({\bf q},\omega)
\frac{{\mbox{\boldmath $\sigma$}}^j \cdot {\rm {\bf q}}_ j}
{{\mbox q}_j^2 +  \mu^2 - \epsilon_j^2 }
\left\{ [{\mbox{\boldmath $\tau$}}^k\times {\mbox{\boldmath
$\tau$}}^j]_3   {\rm {\bf q}} \times
 \left ( {\mbox{\boldmath $\sigma$}}^k \times 
 {\rm {\bf q}}_j  \right ) \right. \right. \nonumber \\
& & \hspace{1.5cm} -  \left. \left. 4
  {\mbox {$\tau$}_3}^j
 {\rm {\bf q}}_k \times {\rm {\bf q}}_j
 \right \}   
+ (k \longleftrightarrow j) \rule[-2ex]{0ex}{4ex}
\right ].
\end{eqnarray*} 

In these equations, ${\bf q}_k$ and ${\bf q}_j$ are the momenta
transferred to each nucleon, $\epsilon_k$ and $\epsilon_j$
the corresponding energies, $\mu$ is the pion mass and 
$f_\pi^2=0.079$ is the pion--nucleon effective coupling constant.
Finally, $F_S$, $F_\pi$ and $F_\Delta$ are the 
form factors of these currents. 

Other MEC mechanisms, such as those corresponding to the
one--rho exchange, the $\pi \rho \gamma$ and the 
$\pi \omega \gamma$, provide
additional terms to the nuclear current operator, all of them 
giving, in general, small contributions to the observables
of interest to us (see Ref.~\cite{Che71} for further details about 
the structure of the different operators).

A common analysis of these current terms can be carried out if
an expansion in powers of $v/c$ (or $1/M$, with $M$  the nucleon
mass) of the nuclear charge and current operators is done.
Following Friar~\cite{Fri83} it is possible to state that both
the IA current as well as MEC are of order $(v/c)^1$, while the
leading term in the charge operator, $\rho^{IA}$, is of order
$(v/c)^0$. Corrections to this term are of order $(v/c)^2$ and
are of relativistic type including MEC pieces. This is one of the
reasons why MEC contributions have been considered extensively in the
current sector and only in a few cases in the charge one. Also
this indicates where to look for identifying the corresponding
effects in this last case: those situations where the IA
contribution vanishes or is very small.

The fact that those MEC contributions to the charge are
relativistic makes that they are not fixed by the
non--relativistic CE above discussed. In this sense, these
charge contributions are model dependent but, fortunately, the
larger effects are produced by the seagull--type term of the pion
exchange current. This is given by~\cite{Neg78,Ris83a}
\begin{eqnarray} 
\label{mecch}
\rho^{\rm MEC}({\bf q},\omega) &=& \sum_{k<j} 
 \left \{  4\pi  \frac{f_\pi^2}{\mu^2} 
\frac{1}{4M}  \left[ F_1^S({\bf q},\omega)
{\mbox{\boldmath $\tau$}}^k \cdot {\mbox{\boldmath $\tau$}}^j 
+ F_1^V({\bf q},\omega)  \tau_z^j \right] \right. \\ 
 && \left. \nonumber \hspace{2.5cm}
{\mbox{\boldmath $\sigma$}}^k \cdot {\bf q}  
\frac{{\mbox{\boldmath $\sigma$}}^j \cdot 
{\rm {\bf q}}_j}{{\mbox q}_j^2 +
 \mu^2 - \epsilon_j^2 }
 + \; \; (k \longleftrightarrow j) \right \} ,
\end{eqnarray} 
where $F_1^S$ and $F_1^V$ are the isoscalar and isovector
nucleon form factors, respectively. Actual calculations have
been performed by using the parametrizations of Ref.~\cite{Hoe76}
for the different nucleon form factors.

Other terms involving the exchange of the $\rho$ and 
$\omega$ mesons and other, such as the $\rho \pi
\gamma$, produce small contributions.
Calculations in the deuteron~\cite{Gar76a} and the
$\alpha$--particle~\cite{Gar76} show that these
additional MEC contributions are only important for very high
$q$--transfer. The pion--exchange term dominates by more than 
one order of magnitude until $q \sim 5$~fm$^{-1}$ 
where the $\rho \pi \gamma$ starts to be the leading one. 
In heavier nuclei the situation is similar~\cite{Gar76} and 
the modifications introduced by the nuclear medium in these nuclei 
only change slightly the results one obtains
with the bare pion exchange term alone~\cite{Ris83a}.

In what follows we analyze the role of the MEC contribution to
the charge operator of Eq.~\ref{mecch} in different electron
scattering processes involving medium and heavy nuclei. Until
now, the cases investigated have been the charge form factors of
different nuclei such as $^{16}$O and
$^{40}$Ca~\cite{Gar76,Ris83} and $^6$Li~\cite{Lod85} and the
charge densities of closed--shell nuclei throughout the periodic
table~\cite{Neg78}. The inclusion of these contributions allow,
in the first case, the description of some of the diffraction
minima in the charge form factor and, in the second, a better
agreement with the experimental data. Nevertheless, it is worth
to say that, in all the cases, the effect is rather small. We
want to analyze if the same conclusion can be drawn when nuclear
excitation is considered.

\section{Quasi--elastic peak}

The first situation we analyze concerns with the QE
peak. In this energy region, the main problem deals with the
longitudinal, $R_L$, and transverse $R_T$ responses, which are
related to the cross--section in the following way:
\[
\frac{d\sigma}{d\Omega'dE'}=  
\sigma_{\mbox{\scriptsize\rm M}} 
    \left[ \frac{q_{\mu}^4}{{\bf q}^4}R_L(q,\omega)+  
    \left(\tan^2\frac{\theta}2-\frac{q_{\mu}^2}{2{\bf q}^2}\right)
    R_T(q,\omega)  \right] .
\]
Here $q^{\mu}=(\omega,{\bf q})$ is the four--momentum transferred
to the nucleus, $\theta$
is the scattering angle and
$\sigma_{\mbox{\scriptsize\rm M}}$ is the Mott cross--section,
\[
 \sigma_{\mbox{\scriptsize\rm M}} = \left(\frac{\alpha\cos(\theta/2)}
 {2E\sin^2(\theta/2)} \right)^2 .
\]
The response functions $R_L$ y $R_T$ are given in terms of the
transition matrix elements of the nuclear charge and current
operators discussed above, between the ground state, $| 0
\rangle$, and final states, $| n \rangle$, of the nucleus:
\begin{eqnarray*}
 R_L(q,\omega) & = & \sum_{n}  \delta(E_n-\omega)
 |\langle n | \rho({\bf q}) | 0 \rangle |^2  \\
 R_T(q,\omega)  & = & \sum_{n} \delta(E_n-\omega)
 |\langle n | {\bf J}_T({\bf q)} | 0 \rangle |^2 .
\end{eqnarray*}

As it is well known, these cross--sections can be described by
means of a simple Fermi gas model in a very good
way~\cite{Mon71}. However, once the separation of the two
responses is performed, it can be realized that neither the
Fermi gas model nor more sophisticated approaches predict the
experimental results: the longitudinal response is usually
overestimated, while the transverse one is
underestimated~\cite{Fro85}.

Different mechanisms have been considered to solve this problem
(2p--2h configurations, final state interactions, relativistic
effects, etc.), but though the situation of the longitudinal
response is more or less understood, the same does not occurs
for the transverse, where the inclusion of the MEC is not
sufficient to bring theory and experiment into full agreement.

What we want to investigate is the role of $\rho^{\rm MEC}$ in 
the longitudinal response. To do that we consider the Fermi gas 
model to describe the nucleus. In this model, the longitudinal 
response can be written as
\[
R_L = R_L^{\rm IA} +  R_L^{\rm IA-MEC} + R_L^{\rm MEC} ,
\]
where the second term (the interference one) gives the larger
contribution of MEC, the last being negligible. The details of
the calculation of this response for one particle--one hole
(1p--1h) final states are given in Appendix~A.

In our calculations we assume symmetric nuclear matter. As shown
in eq.~(A.1) of Appendix~A, the important point is that the
contribution of the interference between the IA and the MEC
contributions to the charge response is negative. The effect of
adding this term to the IA longitudinal response is then to
reduce it, an effect which goes in the direction of reaching the
experimental data.

Now we evaluate the extent of such reduction. First we calculate
the responses for the Fermi momentum of nuclear matter ($k_{\rm
F}=272$~MeV/$c$) for different nuclei of closed--shell nuclei and
for three values of the momentum transfer. To measure the effect 
of the MEC we analyze the relative contribution
\begin{equation}
\displaystyle
\label{rela}
r =\frac{ ( R_L - R_L^{\rm IA} ) }{ R_L^{\rm IA} }
\end{equation}
Table~1 shows the values obtained at the peak position in each
case. It can be seen that the addition of MEC contributions only
produce a small reduction of the IA response, no larger of 3\%.

In any case, it is important to note (see Fig.~1) that the
response $R_L^{\rm IA-MEC}$ (long--dashed lines) presents its
maxima at lower energies than $R_L^{\rm IA}$ (short--dashed
lines) and $R_L$ (solid lines): 40~MeV for $q=300$ and
400~MeV/$c$ and 80~MeV for $q=500$~MeV/$c$. This implies a
certain dependence of these MEC effects with the excitation
energy.

In Fig.~2 we show (solid lines) the results obtained for the
$r$--parameter as a function of $\omega$. The curves correspond
to the different nuclei analyzed and, at the scale of the
figure, are overlapping. Thus, it is apparent that the
effect does not depend on the nuclei. On the other hand, the
contribution of MEC is a small correction to the IA response and
only for small energies and high momentum transfer $r$ reach
values of the order of a 10\%.

The results just quoted have been obtained for $k_{\rm
F}=272$~MeV/$c$. A second aspect of interest is the possible
dependence of the MEC effects with the values of the Fermi
momentum. As it is well known, the Fermi gas model has been used
to describe the QE response of finite nuclei by readjusting the
value of $k_{\rm F}$ to adequate values~\cite{Mon71}. These
values can be obtained by averaging the Fermi momentum with the
density~\cite{Ama94}. 

In Fig.~2 we have plotted also the $r$--parameter calculated for 
$^{12}$C (dashed--dotted lines) and 
$^{40}$Ca (dashed lines), for the three momentum transfer
considered and for $k_{\rm F}=215$ and 235~MeV/$c$,
respectively. This particular values are those provided by the
procedure mentioned above. As we can see, the consideration of
the new $k_{\rm F}$ values reduces the effect of the MEC and
the modification of the IA longitudinal response is, at the peak
energies, negligible in practice. The general trends pointed out
for the nuclear matter Fermi momentum are still valid.

The main conclusion one can draw is that, though a reduction
of the one--body longitudinal response in the QE
peak occurs after adding the MEC contribution to the charge operator, 
it does not produce a sizeable effect. The
smallness of these MEC contributions is due to the fact that the
charge response is dominated by the proton excitations, which
are by far considerably large. As noted in the Introduction,
MEC effects were seen in cases were the IA term is negligible
and this is not the case of the QE region. Besides, the precision 
of the data, clearly worst than those of the charge
distributions quoted in Ref.~\cite{Neg78},
makes not necessary the consideration of these MEC 
effects in the calculation of such responses.

\section{Electroexcitation of bound levels in closed--shell nuclei}

The possibility of considering giant resonances to
look for this MEC effect is not reasonable because of the
difficulties inherent to this energy region in what refers to
the nuclear structure. Besides, from the point of view of the
dominant transitions, the mixing of different multipolarities 
at any energy is similar to that observed in the QE peak and
one can expect the corresponding electromagnetic cross--sections
to be only slightly affected by the presence of this term
in the charge operator.

Low energy transitions offer, at least in principle, much more
opportunities in order to investigate the 
modifications of the charge nuclear operator induced by MEC, because of the
capacity for selecting transitions with given characteristics.
In what follows we focus on these transitions. In order to
minimize at most the uncertitudes associated to the nuclear
structure, we will consider closed--shell nuclei.
 
We study the process of the electroexcitation of low energy
levels. The corresponding cross section is given by~\cite{deF66}
\[
\frac{d\sigma}{d\Omega}=Z^2  
\sigma_{\mbox{\scriptsize\rm M}} \frac{1}{\eta} 
    \left[ \frac{q_{\mu}^4}{{\bf q}^4} |F_L(q)|^2+  
    \left(\tan^2\frac{\theta}2-\frac{q_{\mu}^2}{2{\bf q}^2}\right)
    |F_T(q)|^2  \right] ,
\]
where $Z$ is the nuclear charge, $\eta$ is the recoil factor,
\[
\displaystyle \eta = 1+ \frac{2E}{M_T}{\rm sen}^2 \frac{\theta}{2},
\]
with $M_T$ the target nucleus mass, and $q_\mu=({\bf q},\omega)$
id the four--momentum transfers. The longitudinal,
$|F_L(q)|^2$, and transverse, $|F_T(q)|^2$, 
form factors include the information relative to the nuclear
structure. In the case of closed--shell nuclei and for a transition
between the ground, $|0^+\rangle$, and the 
excited, $|J^\pi \rangle$, states these form factors are:
\begin{eqnarray*}
 |F_L(q)|^2 & = & \frac{4\pi}{Z^2} 
  |\langle J^\pi \| M_J(q) \| 0^+ \rangle |^2  \\
 |F_T(q)|^2 & = & \frac{4\pi}{Z^2} \left\{
|\langle J^\pi \| T_J^{\rm ele}(q) \| 0^+ \rangle |^2 +
|\langle J^\pi \| T_J^{\rm mag}(q) \| 0^+ \rangle |^2  
 \right\} ,
\end{eqnarray*}
where $M_J(q)$ is the Coulomb operator and $T_J^{\rm ele}(q)$ and
$T_J^{\rm mag}(q)$ are the electric and magnetic transverse operators.
These electromagnetic multipole operators are related to the
nuclear charge, $\rho ({\bf r})$, and current, {\bf J}({\bf r}),
in the following way: 
\begin{eqnarray*}
M_{JM} (q) &=& \int d{\bf r}\, j_J(qr)\, Y_{JM}\, (\hat{\bf r})
\rho({\bf r}) \\[.5cm]
T_{JM}^{\rm ele} (q) &=& \frac{1}{q} 
\int d{\bf r}\, \left\{ 
\nabla \times \left[ j_J(qr)\, {\bf Y}_{JJ}^M (\hat{\bf r}) \right]
\right\} \cdot {\bf J}({\bf r}) \\[.5cm]
T_{JM}^{\rm mag} (q) &=& \int d{\bf r}\, 
 j_J(qr)\, {\bf Y}_{JJ}^M (\hat{\bf r}) \cdot {\bf J}({\bf r}) .
\end{eqnarray*}
Here $j_J(qr)$ is a spherical Bessel function, $Y_{JM}
(\hat{\bf r})$ is a spherical harmonic and ${\bf Y}_{JJ}^M
(\hat{\bf r})$ is a vector spherical harmonic. 

Taking into account the model of the nuclear charge given by 
eqs.~(\ref{ch}) and (\ref{mecch}), the longitudinal form factor, 
the one we are interested in, can be written as:
\[
 |F_L(q)|^2 \,=\, 
|t^{\rm IA}_{{\rm C}J}(q)+t^{\rm MEC}_{{\rm C}J}(q)|^2
\]
with
\[
t^a_{{\rm C}J}(q)\,=\, \frac{\sqrt{4\pi}}{Z}
  \langle J^\pi \| M_J^a(q) \| 0^+ \rangle , \,\,\,\,\,
a={\rm IA,MEC}.
\]

\subsection{Shell--model}

First we study the closed--shell nuclei in the framework of the
extreme shell--model, in which the excited levels are described
as 1p--1h states. The reduced electromagnetic matrix elements
$t^a_{{\rm C}J}$ in this approximation are given by eq.~(B.2) in
Appendix~B.

The evaluation of these reduced matrix elements requires to fix
the corresponding configuration space. In this work we have
used a phenomenological Woods--Saxon potential to generate the
single--particle states and energies of the nuclei considered.
This potential is as follows:
\[
   V(r)=-\frac{V_0}{1+e^{(r-R_0)/a_0}}-V_{\rm LS}
	 \frac{\hbar^2}{\mu^2 c^2}
         \frac{1}{r} \frac{d}{dr}
	 \left(\frac{1}{1+e^{(r-R_{\rm LS})/a_{\rm LS}}}\right)
         {\bf l} \cdot {\mbox{\boldmath $\sigma$}}
         +V_{C}(r) ,
\]
where $V_C(r)$ is the Coulomb potential generated by a uniform
spherical distribution with radius $R_c$. The parameters of this
potential are adjusted in order to reproduce the single particle
energies of the levels around the Fermi level and the root mean
squared charge radius of the corresponding closed--shell nuclei.
Table~2 shows the values used in this work.

\subsubsection{$^{16}$O and $^{40}$Ca.} 

The first question we want to analyze is the behaviour of
$\rho^{\rm MEC}$ depending on the isospin of the transition. To
do that we have calculated the longitudinal form factor for the
$^{16}$O and $^{40}$Ca nuclei and the transitions to levels with 
the low excitation energies (see Table~3). 

The fact that in both cases $Z=N$ permits a direct
comparison of the results obtained for protonic and neutronic
transitions with the same quantum numbers. These are drawn in
Fig.~3 and 4. Therein it is apparent that the curves including
the two--body term (solid curves) are considerably modified with
respect to those calculated with $\rho ^{\rm IA}$ (dashed
curves) in the case of transition of neutronic type, even at low
momentum transfers. The effect of the MEC in protonic
transitions are much smaller. This situation is shown in
Table~4, where we give the values of the factor 
\begin{equation}
\displaystyle
r =\frac{  |F_L|^2 - |F_L^{\rm IA}|^2  }{ |F_L^{\rm IA}|^2 }
\label{rela1}
\end{equation}
obtained, in each case, at the scattering maximum with larger strength.
It can be seen that the $\rho ^{\rm MEC}$ term dominates the longitudinal 
form factor in some cases of neutronic character, while the
effects are practically negligible for protonic transitions.

However, these two nuclei are not expected to present purely
neutronic transitions. The similitude between the energies of
the 1p--1h configurations for protons and neutrons suggests
actual excited states composed by a strong mixing of these
configurations.  As a consequence, the large difference in the
strength of $|F_L|$ for both types of transitions (three orders
of magnitude in $^{16}$O and two in $^{40}$Ca) will hide the
effects of the MEC charge operator.

The idea is then to look for nuclei in which there exist
excitations dominated by neutronic 1p--1h configurations. This
occurs for nuclei with neutron excess. In the case of
closed--shell nuclei, $^{48}$Ca is the first one with this
property. In what follows we analyze this nucleus and, in
addition, we study the $^{208}$Pb. The comparison of the results
obtained for both nuclei will provide us a good information on
the question we are discussing.

\subsubsection{$^{48}$Ca} 

In this nucleus, the subshell $1f_{7/2}$ is filled for neutrons
and empty for protons. This situation favours the appearance of
excited states predominantly neutronic at low energy and with
low multipolarity. Table~5 shows the final states of 1p--1h
type that can be built in the extreme shell model approach. As
can be seen, the presence of the multipolarities $1^-$ and $3^-$
for both protons and neutrons with similar energies,
reveals again the large mixing one can expect for these
multipolarities in the actual excited states, and we have not
considered them here.

On the other hand, the multipolarities with parity plus are only
possible for excitations of neutronic type. This is the type of
levels we are interested in. Besides we have considered the
$7^-$ and the $5^-$ levels. In the first case, its high--spin
character reduces the 1p--1h configurations allowed. In the
second one, the expected mixing should maintain a certain
degree of purity because of the energy differences ($\sim
2$~MeV) between the corresponding 1p--1h permitted.

The results obtained for the longitudinal form factors are
plotted in Fig.~5. As in the previous cases, it is evident how
the effect of adding the MEC term is completely negligible in
the case of protonic transitions while appreciable modifications of the
one--body results are observed for transition to some neutronic
final states. 

In order to quantify these effects we have
evaluated the $r$--factor, as given by eq.~(\ref{rela1}), for the
scattering maxima with larger $|F_L|$ in the different
transitions. The results obtained are shown in Table~6. The main
aspect to be noted is that the values for $r$ are now
considerably smaller than for $^{40}$Ca. However, the fact that
some of the actual excitations in this nucleus are dominated by
neutron 1p--1h configurations, makes it still particularly
interesting. Another point to be noted is the fact that the
transition to the $\pi(1f_{7/2},1d_{5/2}^{-1})_{5^-}$ shows a 8\%
effect, a value notably bigger that those obtained for $^{16}$O
and $^{40}$Ca. 

In order to elucidate if some feasible consequence
can be extracted from the results obtained for this nucleus,
more realistic calculations, such as, e.g., RPA, are needed.
Nevertheless, we analyze the case of $^{208}$Pb before going to
this point.

\subsubsection{$^{208}$Pb} 

In this nucleus the levels of interest are the high--spin
states. At low energy the excited states with low angular
momentum show a great collectivity and then the MEC effects
cannot be observed without ambiguities.

High--spin states do not present this problem because of the
reduced possibilities of coupling. As a consequence, the
corresponding wave functions are dominated by only a few (one or
two at most in many cases) 1p--1h configurations and are
ideal for the study we are carrying out.

The states of this type which have been investigated in deep are
those of magnetic character. From its discovery~\cite{Lic78}
they have been considered to analyze some aspects related to the
nuclear structure (partial occupation of single--particle 
levels~\cite{Pan84}, the role of 2p--2h configurations~\cite{Kre80}, 
the effective residual interaction~\cite{Co89}, ...) or,
even, to determine the MEC effects~\cite{Deh85}.

However, there exist also~\cite{Lic79} electric states of this
type, which are those we want to study here. In Table~7 the
configurations giving rise to these states in the shell--model
are shown. As in the case of $^{48}$Ca, the possibilities of
coupling of protonic configurations are strongly restricted. On
the other hand, the proximity in the excitation energies
predicts the mixing of 1p--1h configurations in the realistic
wave functions and this will permit to study situations slightly
different to those seen in the previous cases. In particular, it
is interesting to point out the cases of the $10^+$ states.

Fig.~6 shows the results corresponding to the transitions with
$J \geq 10$. In Table~8 we have
included the values of the $r$--factor at the maxima of $F_L$
with larger strength.
It can be seen how the inclusion of $\rho^{\rm
MEC}$ produce notable effects in some of the neutronic
transitions. Also, it is worth to point out that, again,
we find a protonic transition, the one with final state 
$\pi(1i_{13/2},1h_{11/2}^{-1})_{11^-}$, with a considerably high
$r$--value.

\subsubsection{Analysis of the results}

The results quoted in this section deserve some comments with
respect to the regularities they show and which are of interest.

The first point, above noted, is the considerable difference
between the relative effects of the MEC in the transitions of
protonic and neutronic character. This is obviously due to the
smallnes of the one--body contribution in the neutron case. In
any case, it has been observed some transitions of protonic
character for which $r$ is larger than the 3\% observed in the
quasi--elastic peak.

A second characteristic shown by the results is the reduction of
the MEC effects as the multipolarity grows. This can be observed
in those cases where various multipolarities are accesible to
the same 1p--1h configuration (see the results quoted in
Tables~4, 6 and 8). Then, the advantage that a priori
constitues the consideration of, e.g., the high--spin states in
order both to select those of neutronic character and to
minimize the number of possible 1p--1h configurations
contributing to the wave functions could disappear be
elliminated because of the experimental inability to detect the
effects we are looking for.

Finally, a third aspect which must be pointed out is the strong
variation between the MEC contributions corresponding to the two
possible couplings permitted, for a given multipolarity, by the
two single--particle states with same orbital angular momentum
$l$.  This situation can be observed in different cases. For
example, the transitions to the levels $\nu
(2p,1f_{7/2}^{-1})_{4^+}$ in $^{48}$Ca, (see Table~6 and Fig.~8)
show MEC effects apreciably bigger for the single--particle
state $2p_{3/2}$ than for the $2p_{1/2}$. Same occurs in
$^{208}$Pb (see Table~8 and Fig.~9) with the transitions to the
$\nu (2g,1i_{13/2}^{-1})_{10^+}$ and $\nu
(1j_{15/2},2f^{-1})_{10^+}$, where the relative variation due to
the MEC is larger for the single--particle states $2g_{9/2}$ and
$2f_{7/2}^{-1}$, respectively.

\subsection{RPA}

Once we have analyze the basic aspects of the $\rho^{\rm MEC}$
operator by means of the shell--model, it is interesting to go
deeper and study more realistic situations which permit the
experimental identification of these effects or the
determination of the conditions in which it is necessary or not
to incorporate this operator to the model. To do that, and
taking into account the characteristics of the nuclei we are 
considering, we will investigate the properties of the MEC
contributions to the charge operator in the RPA framework.

This approach represent one of the procedures mostly used to
describe the structure of closed--shell nuclei. 
In this approximation, the nuclear levels are given in terms
of linear combinations of 1p--1h states which are obtained by
acting with the pair p--h creation and anihilation operators
on the correlated vaccum $|C\rangle$:
\begin{equation}
|J M \rangle = \sum_{ph} 
 \left\{ X_J(ph) [ a_p^{\dag} \otimes b_h^{\dag} ]^J_M
- (-1)^{j_p+j_h+M} Y_J(ph) [ b_h \otimes a_p ]^J_{-M}
\right\} | C \rangle .
\end{equation}
Here $a^{\dag}$ ($a$) and $b^{\dag}$ ($b$) are the particle and
hole creation (anihilation) operators, respectively and 
$X$ and $Y$ are the RPA amplitudes. These are calculated by solving
the RPA equations which involve the matrix elements of the residual
interaction.

Our calculations have been performed using a residual
interaction of Landau--Migdal. This is a zero--range interaction
of the form
\begin{eqnarray*}
V_{\rm LM} ({\bf r}_1,{\bf r}_2) &=& C_0 \,\delta({\bf r}_1-{\bf r}_2)
\left\{ f_0^{\rm ex}+ (f_0^{\rm in}-f_0^{\rm ex})\rho({\bf r}_1)
\right.
\\
&& \left.
\, + \, f_0' \,
{\mbox{\boldmath $\tau$}}^1  \cdot {\mbox{\boldmath $\tau$}}^2
\, + \,
g_0\, {\mbox{\boldmath $\sigma$}}^1\cdot {\mbox{\boldmath $\sigma$}}^2 
+\, g_0'\,
  {\mbox{\boldmath $\sigma$}}^1\cdot {\mbox{\boldmath $\sigma$}}^2
  {\mbox{\boldmath $\tau$}}^1  \cdot {\mbox{\boldmath $\tau$}}^2
\right\}
\end{eqnarray*}
where $\rho({\bf r})$ is the nucleon density and
the parameters (see Table~9) have been taken from
Ref.~\cite{Rin78} with the change of $g_0$ from
0.55 to 0, a more realistic value for this type of
interaction~\cite{Co89}. In any case it should be mentioned that the 
channels $g_0$ y $g'_0$ do not affect to much the calculation of
electric states.

In Figs.~10 and 11 we show the spectra obtained in our
calculations for $^{48}$Ca and $^{208}$Pb. Only the levels of
interest for our study have been included for an easier
analysis. As it can be seen, the calculation is
reasonnable for lead, despite the scarcity of data. On the other
hand, the results are not satisfactory for calcium, what indicates 
the necessity of a more adequate interaction, in this case. This
aspect is out of the purpose of this work and we will use the
same interaction in both nuclei.

Once the excited states wave functions are determined, the
reduced matrix elements are evaluated as given by eq.~(B1).
The single--particle wave functions needed for this calculation 
have been generated with the same Woods--Saxon potential used 
in the case of the shell--model.

The results obtained for the $r$--factor for the two nuclei we are
studying are summarized in Tables~10 and 11. 

In the case of $^{48}$Ca (see Tables~6 and 10) it can be stated
that the results are similar to those observed in the
shell--model calculations, though the configuration mixing
produced by the RPA reduces considerably the MEC effects in most
cases. It is worth to point out the transitions to the
$7^-$ at 8.95~MeV and $5^-$ at 8.99~MeV and 11.03~MeV, which are
the only in which MEC effects are above 5\%. In the three cases (see
Fig.~9) the corresponding transverse form factors (right pannels) 
take values of the same order as the longitudinal ones (left pannels), 
what could permit their separation and empirical determination. 
In this figure, solid (dashed) curves have been
obtained with (without) the inclusion of MEC. In the transverse
form factor the model for the current we have considered
includes the seagull and pionic terms
discussed in Sect.~2.

We can conclude that, in this case, the study of $5^-$ levels
and the search of the $7^-$ could permit to go deeper in the 
investigation of the role of
the MEC corrections to the nuclear charge in this nucleus.
However, the relatively unsatisfactory results quoted above with
respect to the residual interaction used, needs of a more
detailed study of this aspect. Work in this direction is in
progress~\cite{Rub96}.

The case of $^{208}$Pb is quite different. Though the RPA calculation
produce a reduction of the relative effects of MEC, as it is
apparent from Figs.~9 and 13, such a reduction is not so
pronounced as in $^{48}$Ca. The numerical values of $r$ shown in
Table~11 favour this conclusion. In particular, the transitions
to the $11^-$ at 6.54~MeV, $13^-$ at 6.55~MeV and $10^+$ at
7.22~MeV states, all of them with a dominant component of
neutronic character, are still showing an appreciable influence
of $\rho^{\rm MEC}$. Besides, the transition to the $11^-$ at
7.24~MeV state maintain the $r$--value found for its dominant
configuration which is of protonic type.

On the other hand, and as it can be seen in Fig.~10, the 
longitudinal (left pannels) and transverse (right pannels) form 
factors for these four cases show similar order of magnitude and 
this makes the separation of both form factors experimentally 
feasible.

The importance of the configuration mixing produced by the RPA
is evident in the results we have just discussed. It can be seen
in Tables~10 and 11 that when the dominant configuration
contributes to the wave function with an amplitude 
$[X_J(ph)+(-1)^J Y_J(ph)]$ far from the value 1 (that is, when
the mixing is stronger), the reduction of the $r$--factor is
considerable. This corroborates the comments made in the case of
the quasi--elastic peak and indicates in a clear way that the
MEC effects we are analyzing will be important only when the
wave function of the excited level is sufficiently pure and,
preferably, of neutronic character. These are the cases in which
the $\rho^{\rm MEC}$ operator must be included in the calculations.

\section{Summary and conclusions}

In this work, the corrections introduced by the MEC in the
nuclear charge operator have been analyzed in nuclear 
electroexcitation processes at different energy regions.

In the quasi--elastic peak, the effect of these contributions produces 
a diminution of the longitudinal response. Though this goes
in the direction of the experimental results, the reduction is of
the order of a 3\% only at the peak energy, and then it is insufficient 
to provide the agreement with the data. Taking into account the 
incertitudes introduced in the longitudinal response by the nucleon form
factors and the experimental error, we can conclude that the inclusion
of $\rho^{\rm MEC}$ in the calculations at these energies is irrelevant 
at present.

At low energy, MEC effects in the longitudinal form factor result 
to be relatively important in some cases. Calculations performed in the 
extreme shell--model point out the strong asymmetry of such contributions
depending on the isospin of the transition: they are considerably larger 
in case of neutronic excitations.

On the other hand, the results obtained with calculations of RPA type
reduce the MEC effects due to the configuration mixing generated. 
Nevertheless, it is possible to found situations in which the 
consideration of the $\rho^{\rm MEC}$ operator is mandatory.

Some of the observed aspects could be amplified if processes including
polarization degrees of freedom and involving closed--shell nuclei
$\pm 1$ nucleon are used. In previous 
works~\cite{Ama94a,Amayo} it has been shown how the selection of the 
target polarization axis together to an adequate kinematics allow the
appearance of big effects due to the MEC contributions to the current
operator. In view of the fact that, as seen above, the effects in the
charge (longitudinal) channel are of the same order as those in the 
tranverse one, the mentioned processes could be specially sensible to 
the presence of $\rho^{\rm MEC}$.

In order to finish, it is worth to note that the previous discussions 
have a fundamental point of incertitude. The fact that the most relevant 
aspects are associated to neutronic transitions makes the electric form
factor of the neutron to be one of the basic ingredients. However, the
poor experimental determination of this form factor give rise to large 
differences between the different parametrizations available, what 
introduces ambiguities in the calculations of difficult evaluation. The
reduction of the experimental error for the neutron electric form factor 
is then a basic point to bring our conclusions to a feasible practical 
level. In this sense, the experimental program to be developped at CEBAF
is of maximum interest. On the oppposite side, the careful selection of
some particular transitions (such as, e.g. that to the $13^-$ state in
$^{208}$Pb) and the experimental separation of the two form factors could
provide some alternative insight in the problem of the values of this 
form factor.

\appendix
\section*{Appendix A. MEC contribution to the charge response in
the Fermi gas model}

In this appendix we describe the key points for the calculation
of the interferece response between the $\rho^{\rm MEC}$
operator and the IA charge one in the Fermi gas model we have
considered to analyze the quasi--elastic peak region.

Assuming final states of 1p--1h type, this interferencen term can
be written as:
\begin{eqnarray*}
 R_L^{\rm IA-MEC}({\bf q},\omega)& =& 2 {\rm Re}
\sum_{\rm ph} \delta(\omega+\omega_h-\omega_p)
	            \theta(k_{\rm F}-h)\theta(p-k_{\rm F}) \\ \nonumber
  && \hspace{2cm}
   \langle {\rm ph}^{-1}|\rho^{\rm IA}({\bf q})| F \rangle
   \langle {\rm ph}^{-1}|\rho^{\rm MEC}({\bf q})| F \rangle ^* ,
\end{eqnarray*}
where $k_{\rm F}$ is the Fermi momentum.
Here we have used k to label the single--particle state 
$|{\bf k} \omega_k s_k t_k \rangle$ which is
characterized by a momemtum {\bf k}, with $k=|{\bf k}|$,
an energy $\omega_k$ and third spin and isospin components
$s_k$ and $t_k$, respectively.

Taking into account Eqs.~(\ref{ch}) and (\ref{mecch}), a simple
calculation gives for the response of interest:
\begin{eqnarray*}
 R_L^{\rm IA-MEC}({\bf q},\omega)& =& 4 \pi  \frac{f_\pi^2}{\mu^2} 
\sum_{t_k t_h} \frac{G_{\rm E}^h({\bf q},\omega)}{M_h} 
B^*_{hkkh} {\cal R}_{t_k t_h} \\ \nonumber
&=&-2 \pi  \frac{f_\pi^2}{\mu^2} 
\sum_{t_k t_h} \frac{G_{\rm E}^h({\bf q},\omega)}{M_h} 
B^*_{hkkh} {\cal R}^{\rm MS}_{t_k t_h}({\bf q},\omega)
\end{eqnarray*}
where k runs over all occupied single--particle states,
\[ 
B_{hkkh}=\langle \left(\frac12 t_h \right) 
                 \left(\frac12 t_k \right) | 
F_1^S(q^2) 
{\mbox{\boldmath $\tau$}}^1 \cdot {\mbox{\boldmath $\tau$}}^2 
+ F_1^V(q^2)  \tau_z^2 | 
                 \left(\frac12 t_k \right) 
                 \left(\frac12 t_h \right)  \rangle
\]
and $\cal{R}^{\rm MS}$ represents the reduced transverse response 
corresponding to the interference between the magnetization and 
seagull currents (see Ref.~\cite{Ama94} for details).
For symmetric nuclear matter, the interference response
simplifies to:
$$
R_L^{\rm IA-MEC}({\bf q},\omega)\, =\, -2 \pi \frac{f_\pi^2}{\mu^2} 
\frac{1}{M} 
  \left\{
3\left[G_{\rm E}^{\rm P}({\bf q},\omega)+
       G_{\rm E}^{\rm N}({\bf q},\omega)\right]
F_1^S({\bf q},\omega)  \right. \eqno({\rm A}.1)$$
$$\left. \hspace{7.4cm}
+ \,\left[G_{\rm E}^{\rm P}({\bf q},\omega)-
      G_{\rm E}^{\rm N}({\bf q},\omega)\right]
F_1^V({\bf q},\omega)
\right\} {\cal R}^{\rm MS}({\bf q},\omega) .
$$

\appendix
\section*{Appendix B. MEC contribution to the charge form factor
in the RPA and shell-model approach}

In this appendix we give the expressions of the reduced
matrix elements of the Coulomb operator which enter in
the calculation of the longitudinal form factor. 

In describing closed--shell nuclei in the RPA framework, the nuclear 
wave functions are given by Eq.~(5).
With this definition, and after a cumbersome calculation
involving angular momentum algebra, the reduced matrix elements
involved in the longitudinal form factor can be written as follows:
$$
t^a_{{\rm C}J}(q)\, = \, 
\frac{1}{Z} \sum_{ph} [X^*_J(ph)+(-1)^J Y^*_J(ph)]
(-1)^{j_p-1/2} \xi (l_p+l_h+J)  \eqno({\rm B}.1)$$
\vspace{-.5cm}
$$ \widehat{j}_p \widehat{j}_h \widehat{J} 
 \left( \begin{array}{ccc} {j_p}&{j_h}&{J}\\ 
                           {1 \over 2}&-{1 \over 2}&0 
        \end{array} \right)
\int dx x^2 j_J(qx) {\cal C}^a_{ph;J} (x) ,
$$
where $\xi (\lambda)=1$ or 0 according to $\lambda$ be even or odd,
$\widehat{\lambda}=\sqrt{2\lambda+1}$ and the functions ${\cal
C}^a_{ph;J} (x)$ are:
\begin{eqnarray*}
\label{cia}
{\cal C}^{\rm IA}_{ph;J} (x) & = & G_{\rm E}^{ph}(q) R_p(x) R_h(x)\\
\label{cmec}
{\cal C}^{\rm MEC}_{ph;J} (x) & = & \frac{1}{4M_{ph}} 
\frac{f_\pi^2}{m_\pi^2} \sum_{i} \widehat{j}_{i}^2 {\cal T}_{iphi} 
\sum_{L} \widehat{L}^2 
\left( \frac{d}{dx}+\frac{\kappa_p+\kappa_h+2\kappa_i+2}{x}\right)\\
&& \nonumber \left\{ \xi(l_p+l_{i}+L+1) 
 \left( \begin{array}{ccc} {j_p}&{j_{i}}&{L}\\ 
                           {1 \over 2}&-{1 \over 2}&{0} 
        \end{array} \right)^2 
I_{pi}(L,x) R_{i}(x) R_h(x) \, + \, (p \longleftrightarrow h)
\right\} .
\end{eqnarray*}
Here $i$ runs over all the single nucleon states below the Fermi
level, $\kappa_\alpha=(l_\alpha-j_\alpha)(2j_\alpha+1)$, 
\begin{eqnarray*}
{\cal T}_{iphi}&=& F_1^S(q) [2(\delta_{ph,\nu}\delta_{i,\pi} 
      +\delta_{ph,\pi}\delta_{i,\nu})+
      \delta_{phi,\nu}+\delta_{phi,\pi}] \nonumber \\
      &+& F_1^V(q)[\delta_{phi,\nu}-\delta_{phi,\pi}]
\end{eqnarray*}
and
\begin{eqnarray*}
\displaystyle I_{\alpha\beta}(L,x) & =& \int_0^\infty dr r^2 
 \left\{ \left( {d \over dr}+
     {\kappa_\alpha + \kappa_\beta +2 \over r}
      \right)  R_\alpha(r) R_\beta(r)
        \right\} \nonumber \\ &&
\displaystyle \int_0^\infty dk k^2 {2 \over \pi} v_\pi(k) 
j_L(kr)j_L(kx) .
\end{eqnarray*}

We are also interested in describing the nuclear states in the
extreme shell--model. In this case, the excited levels a of
1p--1h character and are given as:
\[
|J M \rangle = [ a_p^{\dag} \otimes b_h^{\dag} ]^J_M | 0 \rangle ,
\]
with $| 0 \rangle$ the Hartree--Fock vacuum. In this approach, the 
reduced matrix element of eq.~(B.1) simplifies to
$$
t^a_{{\rm C}J}(q)\, =\, \frac{1}{Z} (-1)^{j_p-1/2} 
\xi (l_p+l_h+J) \widehat{j}_p \widehat{j}_h \widehat{J} 
 \left( \begin{array}{ccc} {j_p}&{j_h}&{J}\\ 
                           {1 \over 2}&-{1 \over 2}&0 
        \end{array} \right)
$$
$$
\hspace{-3cm}
\int dx x^2 j_J(qx) {\cal C}^a_{ph;J} (x) , \eqno({\rm B}.2)
$$

\newpage

\noindent
\section*{Tables}

\vspace{1cm}

\begin{center} 
TABLE 1 \\
{\sl Values of the relative effect of the MEC on the charge
response (see Eq.~(\ref{rela})) at the peak positions, for
different nuclei. The nuclear matter value for the Fermi
momentum, $k_{\rm F}=272$~MeV/$c$, has been used.} \\[.5cm]
\begin{tabular}{ccrrrrrr}
\hline\hline
 & & \multicolumn{6}{c}{$r$ [\%]} \\
\hline
$q$ [MeV/$c$] & $\omega_{\rm max}$ [MeV] & $^{12}$C & $^{16}$O 
& $^{40}$Ca & $^{48}$Ca & $^{90}$Zr &  $^{208}$Pb \\
\hline
300 & 50  & \multicolumn{3}{c}{-2.07} & -2.09 & -2.08 & -2.09 \\
400 & 90  & \multicolumn{3}{c}{-2.75} & -2.77 & -2.77 & -2.78 \\
500 & 150 & \multicolumn{3}{c}{-2.81} & -2.83 & -2.82 & -2.84 \\
\hline\hline
\end{tabular}
\end{center}

\vspace{2cm}

\begin{center} 
TABLE 2 \\
{\sl Parameters of the Woods--Saxon potential used in this work.
The values for $^{12}$C, $^{16}$O and $^{40}$Ca have been taken
from Ref.~\cite{Ama94a}, those of $^{48}$Ca
from~\cite{Ama92} and those of $^{208}$Pb from~\cite{Rin78}.} \\[.5cm]
\begin{tabular}{ccccccccc} \hline\hline
 & & $V_0$ & $R_0$ & $a_0$ & $V_{\rm LS}$ & 
     $R_{\rm LS}$ & $a_{\rm LS}$ & $R_{\rm c}$ \\
 & & [MeV] & [fm]  & [fm]  & [MeV]    & [fm]     & [fm] & [fm] \\ 
\hline \rule{0cm}{.5cm}
$^{12}$C & $\pi$ & 62.0 & 2.86 & 0.57 & ~3.20 & 2.86 & 0.57 & 2.86 \\
         & $\nu$ & 60.5 & 2.86 & 0.57 & ~3.15 & 2.86 & 0.57 & \\ 
\hline \rule{0cm}{.5cm}
$^{16}$O & $\pi$ & 52.5 & 3.20 & 0.53 & ~7.00 & 3.20 & 0.53 & 3.20 \\
         & $\nu$ & 52.5 & 3.20 & 0.53 & ~6.54 & 3.20 & 0.53 & \\ 
\hline \rule{0cm}{.5cm}
$^{40}$Ca & $\pi$ & 57.6 & 4.10 & 0.53 & 11.11 & 4.10 & 0.53 & 4.10 \\
          & $\nu$ & 55.0 & 4.10 & 0.53 & ~8.50 & 4.10 & 0.53 & \\ 
\hline \rule{0cm}{.5cm}
$^{48}$Ca & $\pi$ & 59.5 & 4.36 & 0.53 & ~8.56 & 4.36 & 0.53 & 4.54 \\
          & $\nu$ & 50.0 & 4.36 & 0.53 & ~7.72 & 4.36 & 0.53 & \\ 
\hline \rule{0cm}{.5cm}
$^{208}$Pb & $\pi$ & 60.4 & 7.46 & 0.79 & ~6.75 & 7.20 & 0.59 & 7.41 \\
           & $\nu$ & 44.3 & 7.46 & 0.66 & ~6.08 & 6.96 & 0.64 & \\ 
\hline\hline
\end{tabular}
\end{center}

\newpage

\begin{center} 
TABLE 3 \\
{\sl Final states of 1p--1h type considered in $^{16}$O and
$^{40}$Ca nuclei to analyze the MEC effect in $|F_L(q)|^2$. The
corresponding energies and the possible multipolarities are 
given.} \\[.5cm]
\begin{tabular}{rcccccccc} \hline\hline
Nucleus & Final state & $E_\pi$ & $E_\nu$ & \multicolumn{5}{c}{$J^\pi$} \\
       &  (1p--1h)  &  [MeV]  &  [MeV]  & &&&& \\
\hline
         & $(1d_{5/2},1p_{1/2}^{-1})$  & 11.52 & 11.52 & &~~~&$3^-$&~~~& \\ 
$^{16}$O & $(2s_{1/2},1p_{1/2}^{-1})$  & 12.01 & 12.40 & $1^-$&&&& \\ 
         & $(1d_{3/2},1p_{1/2}^{-1})$  & 15.11 & 15.36 & $1^-$&&&& \\ 
\hline
 & $(1f_{7/2},1d_{3/2}^{-1})$  & ~7.24 & ~7.27 & &&$3^-$&&$5^-$ \\ 
$^{40}$Ca 
 & $(1f_{7/2},2s_{1/2}^{-1})$  & ~9.77 & ~9.74 & && $3^-$&& \\ 
 & $(1f_{7/2},1d_{5/2}^{-1})$  & 11.18 & 13.24 & $1^-$&&$3^-$&&$5^-$ \\ 
\hline\hline
\end{tabular}
\end{center}

\vspace{2cm}

\begin{center} 
TABLE 4 \\
{\sl Values of the $r$--factor (see eq.~(\ref{rela1})) for the
different transitions considered in $^{16}$O and $^{40}$Ca,
calculated at the scattering maximum with larger strength in
each case.} \\[.5cm]
\begin{tabular}{rccrr} \hline\hline
Nucleus & Final state & $J^\pi$ & \multicolumn{2}{c}{$r$ [\%]} \\
       &  (1p--1h)  &         & {$\pi$} & {$\nu$}  \\
\hline
$^{16}$O & $(1d_{5/2},1p_{1/2}^{-1})$ & $3^-$ &  1.5 & ~11.4 \\ 
         & $(2s_{1/2},1p_{1/2}^{-1})$ & $1^-$ &  0.0 & -13.5 \\ 
         & $(1d_{3/2},1p_{1/2}^{-1})$ & $1^-$ &  1.0 & ~49.9 \\ 
\hline
$^{40}$Ca & $(1f_{7/2},1d_{3/2}^{-1})$ & $3^-$ & 1.4 & ~17.6 \\
          &                            & $5^-$ & 1.7 & ~10.2 \\ 
          & $(1f_{7/2},2s_{1/2}^{-1})$ & $3^-$ & 1.7 & ~25.0 \\ 
          & $(1f_{7/2},1d_{5/2}^{-1})$ & $1^-$ & 5.8 & 343.2 \\
          &                            & $3^-$ & 6.2 & 111.1 \\
          &                            & $5^-$ & 6.5 & ~64.6 \\ 
\hline\hline
\end{tabular}
\end{center}

\newpage

\vspace*{2cm}

\begin{center} 
TABLE 5 \\
{\sl Same as in Table~3 but for $^{48}$Ca.} \\[.5cm]
\begin{tabular}{lcccccccc} \hline\hline
Final state & Energy & \multicolumn{7}{c}{$J^\pi$} \\
(1p--1h)& [MeV] & & & & & & & \\
\hline
$\pi(1f_{7/2},2s_{1/2}^{-1})$ & ~6.18 & & & $3^-$ & & & & \\ 
$\pi(1f_{7/2},1d_{3/2}^{-1})$ & ~6.54 & & & $3^-$ & & $5^-$ & & \\ 
$\pi(2p_{3/2},2s_{1/2}^{-1})$ & ~9.26 & $1^-$ & & & & & &\\ 
$\pi(2p_{3/2},1d_{3/2}^{-1})$ & ~9.62 & $1^-$ & & $3^-$ & & & & \\
$\pi(2p_{1/2},2s_{1/2}^{-1})$ & ~9.69 & $1^-$ & & & & & &\\ 
$\pi(1f_{5/2},2s_{1/2}^{-1})$ & ~9.99 & & & $3^-$ & & & &\\ 
$\pi(2p_{1/2},1d_{3/2}^{-1})$ & 10.05 & $1^-$ & & & & & & \\ 
$\pi(1f_{5/2},1d_{3/2}^{-1})$ & 10.35 & $1^-$ & & $3^-$ & & & & \\
$\pi(1f_{7/2},1d_{5/2}^{-1})$ & 10.78 & $1^-$ & & $3^-$ & & $5^-$ & &\\ 
\hline
$\nu(2p_{3/2},1f_{7/2}^{-1})$ & ~4.80 & & $2^+$ & & $4^+$ & & & \\ 
$\nu(2p_{1/2},1f_{7/2}^{-1})$ & ~6.83 & & & & $4^+$ & & &\\
$\nu(2p_{3/2},1d_{3/2}^{-1})$ & ~7.38 & $1^-$ & & $3^-$ & & & & \\ 
$\nu(2p_{3/2},2s_{1/2}^{-1})$ & ~7.40 & $1^-$ & & & & & &\\ 
$\nu(1f_{5/2},1f_{7/2}^{-1})$ & ~8.39 & &$2^+$&&$4^+$&&$6^+$&\\
$\nu(1g_{9/2},1f_{7/2}^{-1})$ & ~8.81 & $1^-$&&$3^-$&&$5^-$&&$7^-$\\
$\nu(2p_{1/2},1d_{3/2}^{-1})$ & ~9.41 & $1^-$& & & & & & \\ 
$\nu(2p_{1/2},2s_{1/2}^{-1})$ & ~9.43 & $1^-$& & & & & & \\ 
$\nu(1f_{5/2},1d_{3/2}^{-1})$ & 10.97 & $1^-$&&$3^-$&&&& \\ 
$\nu(1f_{5/2},2s_{1/2}^{-1})$ & 10.99 &&& $3^-$&&&& \\
\hline\hline
\end{tabular}
\end{center}

\newpage

\begin{center} 
TABLE 6 \\
{\sl Same as in Table~4 but for $^{48}$Ca.}\\[.5cm]
\begin{tabular}{ccr} \hline\hline
Final state & $J^\pi$ & $r$ \\
 (1p--1h)  &         & {[\%]}  \\
\hline
$\pi(1f_{7/2},1d_{3/2}^{-1})$ & $5^-$ & ~2.5  \\ 
$\pi(1f_{7/2},1d_{5/2}^{-1})$ & $5^-$ & ~8.0  \\ 
\hline
$\nu(2p_{3/2},1f_{7/2}^{-1})$ & $2^+$ & 55.6 \\
                              & $4^+$ & 44.1 \\ 
$\nu(2p_{1/2},1f_{7/2}^{-1})$ & $4^+$ & -8.0 \\
$\nu(1f_{5/2},1f_{7/2}^{-1})$ & $2^+$ & ~8.9 \\
                              & $4^+$ & ~3.6 \\
                              & $6^+$ & ~3.2 \\
$\nu(1g_{9/2},1f_{7/2}^{-1})$ & $5^-$ & 56.3\\
                              & $7^-$ & 37.3 \\
\hline\hline
\end{tabular}
\end{center}

\vspace{2cm}

\begin{center} 
TABLE 7 \\
{\sl Same as in Table~3 but for $^{208}$Pb.}\\[.5cm]
\begin{tabular}{lcccccccc} \hline\hline
Final state& Energy & \multicolumn{7}{c}{$J^\pi$} \\
(1p--1h)& [MeV] & & & & & & & \\
\hline
$\pi(1h_{9/2},1h_{11/2}^{-1})$  & 5.57 & & $8^+$ & & $10^+$ & & & \\ 
$\pi(2f_{7/2},1h_{11/2}^{-1})$  & 6.47 & & $8^+$ & & & & &  \\
$\pi(1i_{13/2},1h_{11/2}^{-1})$ & 7.18 & $7^-$ & & $9^-$ & & $11^-$ & &\\ 
$\pi(1i_{13/2},2d_{5/2}^{-1})$  & 7.52 & & $8^+$ & & & & &  \\
\hline
$\nu(2g_{9/2},1i_{13/2}^{-1})$  & 5.07 & & $8^+$ & & $10^+$ & & & \\ 
$\nu(1j_{15/2},2f_{5/2}^{-1})$  & 5.43 & & $8^+$ & & $10^+$ & & &\\
$\nu(1j_{15/2},3p_{3/2}^{-1})$  & 5.75 & & $8^+$ & & & & & \\ 
$\nu(1i_{11/2},1i_{13/2}^{-1})$ & 5.85 & & $8^+$ & & $10^+$ & &$12^+$ & \\ 
$\nu(1j_{15/2},1i_{13/2}^{-1})$ & 6.49 & $7^-$ & & $9^-$ & &
$11^-$ & & $13^-$ \\
$\nu(1i_{11/2},2f_{7/2}^{-1})$  & 6.55 & $7^-$&&$9^-$&&&&\\
$\nu(3d_{5/2},1i_{13/2}^{-1})$  & 6.63 & &$8^+$ & & & & & \\ 
$\nu(2g_{9/2},1h_{9/2}^{-1})$   & 6.84 & $7^-$& &$9^-$ & & & & \\ 
$\nu(1j_{15/2},2f_{7/2}^{-1})$  & 7.20 &&$8^+$&&$10^+$&&& \\ 
$\nu(2g_{7/2},1i_{13/2}^{-1})$  & 7.56 &&$8^+$&&$10^+$&&& \\
$\nu(1i_{11/2},1h_{9/2}^{-1})$  & 7.62 &$7^-$&&$9^-$&&&& \\ 
\hline\hline
\end{tabular}
\end{center}

\newpage

\begin{center} 
TABLE 8 \\
{\sl Same as in Table~4 but for $^{208}$Pb.}\\[.5cm]
\begin{tabular}{ccr} \hline\hline
Transition & $J^\pi$ & {$r$} \\
 (1p--1h)  &         & {[\%]}  \\
\hline
$\pi(1h_{9/2},1h_{11/2}^{-1})$  & $10^+$ & ~1.5  \\ 
$\pi(1i_{13/2},1h_{11/2}^{-1})$ & $11^-$ & ~7.3  \\ 
\hline
$\nu(2g_{9/2},1i_{13/2}^{-1})$  & $10^+$ & 40.3  \\ 
$\nu(1j_{15/2},2f_{5/2}^{-1})$  & $10^+$ & ~4.8 \\
$\nu(1i_{11/2},1i_{13/2}^{-1})$ & $10^+$ & ~4.5 \\
                                & $12^+$ & ~3.8 \\ 
$\nu(1j_{15/2},1i_{13/2}^{-1})$ & $11^-$ & 49.0  \\ 
                                & $13^-$ & 39.4 \\
$\nu(1j_{15/2},2f_{7/2}^{-1})$  & $10^+$ & 18.7 \\ 
$\nu(2g_{7/2},1i_{13/2}^{-1})$  & $10^+$ & ~0.1 \\
\hline\hline
\end{tabular}
\end{center}

\vspace{2cm}

\begin{center} 
TABLE 9 \\
{\sl Parameters of the Landau--Migdal residual interaction 
used in our RPA calculation.}\\[.5cm]
\begin{tabular}{cccccc} \hline\hline
$C_0$ [MeV fm$^3$] & $f_0^{\rm in}$ & $f_0^{\rm ex}$ & $f_0'$ 
& $g_0$ & $g_0'$\\
\hline
386.04 & 0.20 & -2.45 & 1.5 & 0.0 & 0.70 \\ \hline \hline
\end{tabular}
\end{center}

\newpage

\begin{center} 
TABLE 10 \\
{\sl Same as in Table~4 but for the
different RPA transitions considered in $^{48}$Ca. The dominant
1p--1h configuration and the amplitude of its contribution to the
form factor are also given in each case.}\\[.5cm]
\begin{tabular}{ccrcc} \hline\hline
Energy & $J^\pi$ & {$r$} & dominant & $[X_J(ph)+(-1)^J Y_J(ph)]$\\
\mbox{ [Mev]}  &    &{[\%]} & (1p--1h) & \\
\hline
~4.81 & $4^+$ & ~~4.1 & $\nu(2p_{3/2},1f_{7/2}^{-1})$ & 1.002 \\ 
~4.87 & $2^+$ & ~~3.6 & $\nu(2p_{3/2},1f_{7/2}^{-1})$ & 1.031 \\
~6.95 & $4^+$ & ~~4.1 & $\nu(2p_{1/2},1f_{7/2}^{-1})$ & 0.973 \\
~7.03 & $5^-$ & ~~3.5 & $\pi(1f_{7/2},1d_{3/2}^{-1})$ & 0.863 \\ 
~8.69 & $4^+$ & ~~4.7 & $\nu(1f_{5/2},1f_{7/2}^{-1})$ & 0.961 \\
~8.87 & $2^+$ & ~~4.4 & $\nu(1f_{5/2},1f_{7/2}^{-1})$ & 0.965 \\
~8.95 & $7^-$ & ~20.3 & $\nu(1g_{9/2},1f_{7/2}^{-1})$ & 0.992 \\
~8.99 & $5^-$ & -16.1 & $\nu(1g_{9/2},1f_{7/2}^{-1})$ & 0.850 \\
~9.31 & $6^+$ & ~~4.3 & $\nu(1f_{5/2},1f_{7/2}^{-1})$ & 0.976 \\
11.03 & $5^-$ & ~~7.0 & $\pi(1f_{7/2},1d_{5/2}^{-1})$ & 0.940 \\ 
\hline\hline
\end{tabular}
\end{center}

\vspace{2cm}

\begin{center} 
TABLE 11 \\
{\sl Same as in Table~10 but for $^{208}$Pb.}\\[.5cm]
\begin{tabular}{ccrcc} \hline\hline
Energy & $J^\pi$ & {$r$} & dominant& $[X_J(ph)+(-1)^J Y_J(ph)]$\\
\mbox{ [Mev]}  &    &{[\%]} & (1p--1h)  & \\
\hline
5.10 & $10^+$ & ~2.6 & $\nu(2g_{9/2},1i_{13/2}^{-1})$ & 0.996 \\ 
5.51 & $10^+$ & ~2.2 & $\nu(1j_{15/2},2f_{5/2}^{-1})$ & 0.795 \\
5.75 & $10^+$ & ~1.8 & $\nu(1i_{11/2},1i_{13/2}^{-1})$ & 0.592 \\ 
6.49 & $12^+$ & ~3.5 & $\nu(1i_{11/2},1i_{13/2}^{-1})$ & 1.003 \\ 
6.54 & $11^-$ & 14.5 & $\nu(1j_{15/2},1i_{13/2}^{-1})$ & 0.984 \\ 
6.55 & $13^-$ & 37.5 & $\nu(1j_{15/2},1i_{13/2}^{-1})$ & 0.995 \\
6.57 & $10^+$ & ~1.4 & $\pi(1h_{9/2},1h_{11/2}^{-1})$ & 0.723 \\
7.22 & $10^+$ & 21.4 & $\nu(1j_{15/2},2f_{7/2}^{-1})$ & 0.993 \\ 
7.24 & $11^-$ & ~7.2 & $\pi(1i_{13/2},1h_{11/2}^{-1})$ & 0.983 \\ 
7.77 & $10^+$ & ~3.9 & $\nu(2g_{7/2},1i_{13/2}^{-1})$ & 0.987 \\ 
\hline\hline
\end{tabular}
\end{center}


\newpage

\noindent
\section*{Figure Captions}

\noindent
FIGURE 1 \\
{\sl Longitudinal response as a function of the excitation
energy $\omega$ for the $^{12}$C (left) and ${40}$Ca (right) and
for the three momentun transfers considered. Short--dashed curves 
represent the results obtained with $\rho^{\rm IA}$,
long--dashed curves correspond to the interference responses
$R_L^{\rm int}$ and solid curves give the total response.}

\noindent
FIGURE 2 \\
{\sl $r$--factor (see eq.~(\ref{rela})) as a function of the energy 
$\omega$. With solid lines we show the values obtained for 
$k_{\rm F}=272$~MeV/$c$ and for the
different nuclei considered. The curves are overlapping at the
scale of the figure. Dashed--dotted curves
represent the results corresponding to the average values
$k_{\rm F}=215$~MeV/$c$ for $^{12}$C. The same but for $^{40}$Ca 
and $k_{\rm F}=235$~MeV/$c$ are plotted with dashed curves.}

\noindent
FIGURE 3 \\
{\sl Longitudinal form factor for the electroexcitation
of the 1p--1h levels in $^{16}$O included in Table~3.
Dashed curves correspond to the calculations performed with the
one--body piece of the charge density. Solid lines also include
the MEC term.}

\noindent
FIGURE 4 \\
{\sl Same as in Fig.~3 but for $^{40}$Ca.}

\noindent
FIGURE 5 \\
{\sl Same as in Fig.~3 but for the transitions in $^{48}$Ca
shown in Table~5.}

\noindent
FIGURE 6 \\
{\sl Same as in Fig.~3 but for the transitions in $^{208}$Pb
shown in Table~7.}

\noindent
FIGURE 7 \\
{\sl Excitation spectrum of $^{48}$Ca obtained with the RPA calculation
described in the text (central column). The experimental results
of Ref.~\cite{Wis85} (right column) and the levels corresponding
to the shell--model (left column) are also included.}

\noindent
FIGURE 8 \\
{\sl Same as in Fig.~7 but for $^{208}$Pb. The experimental data
are from Ref.~\cite{Lic79}.}

\noindent
FIGURE 9 \\
{\sl Longitudinal (left) and transverse (right) form factor for
the electroexcitation of the RPA levels of $^{48}$Ca in which
the $r$--factor is above 5\%. Dashed (solid) curves correspond
to the calculations performed without (with) the MEC pieces of
the charge and current densities, respectively.}

\noindent
FIGURE 10 \\
{\sl Same as in Fig.~9 but for $^{208}$Pb.}

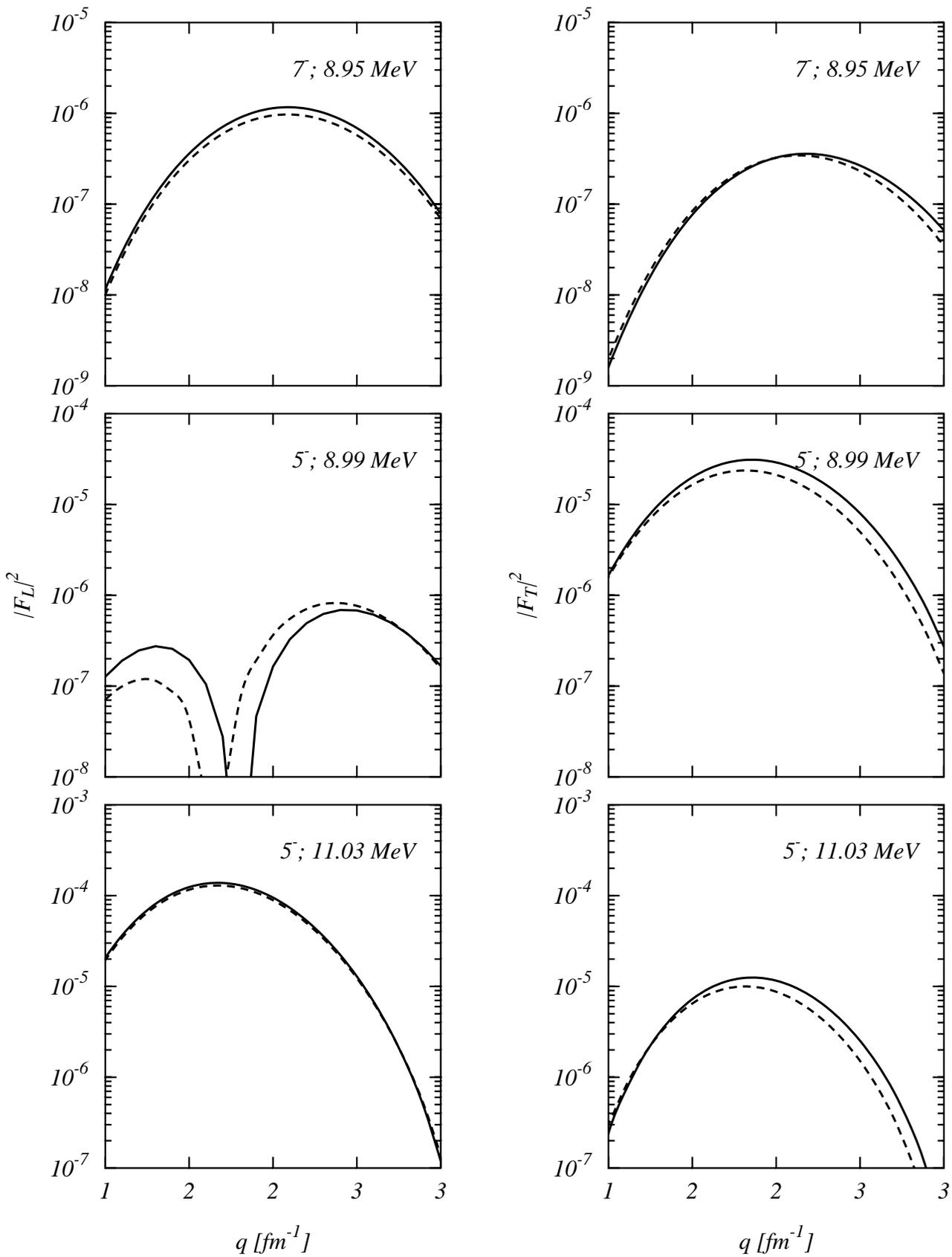
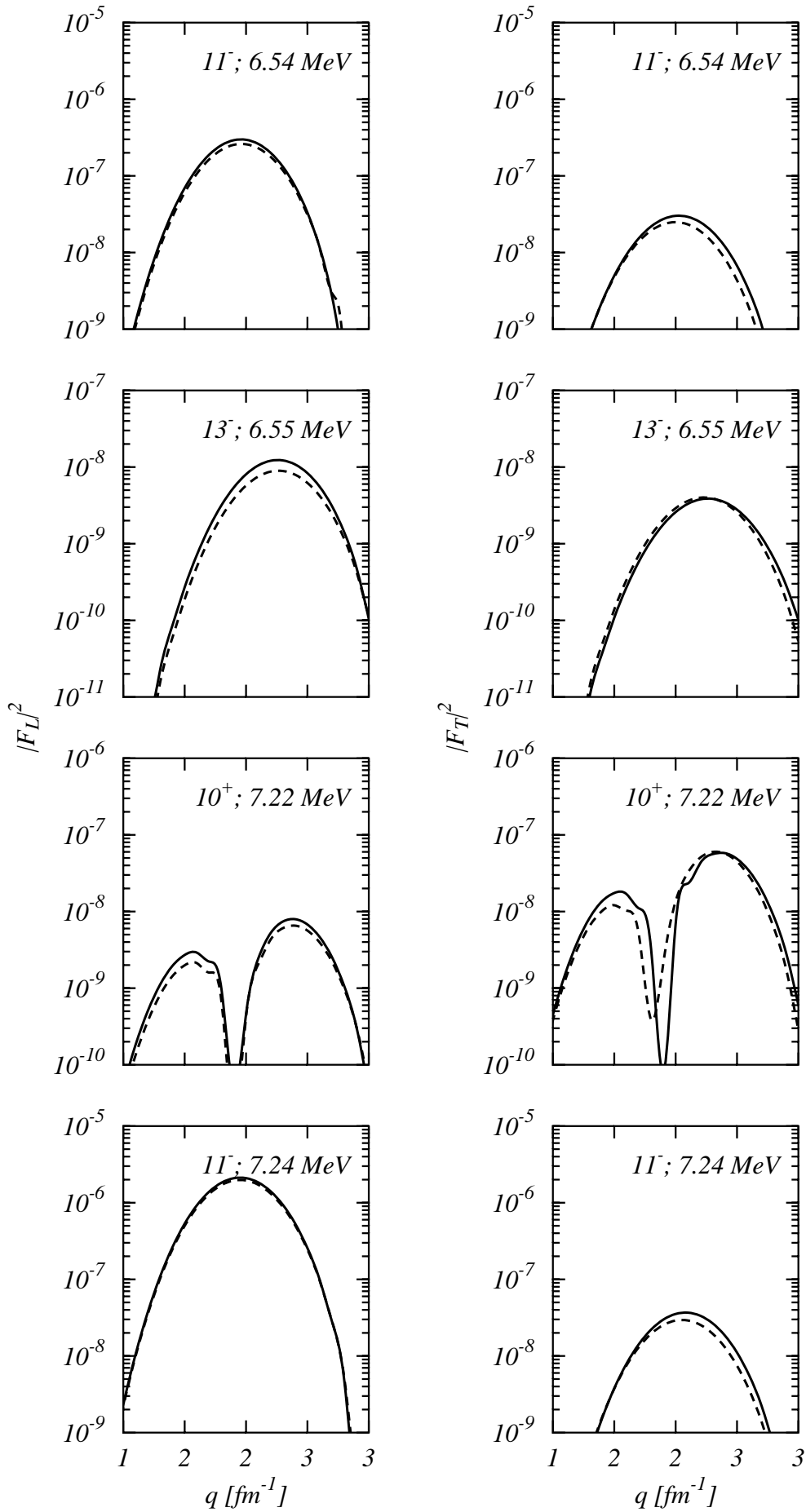
\end{document}